\documentclass[twocolumn,trackchanges]{aastex7}

\usepackage{graphicx}% Include figure files
\usepackage{amsmath}
\usepackage{longtable} % Add this to your preamble
\usepackage{rotating} % Keep this for landscape if needed, but let's try portrait first.
\usepackage{natbib} 
\usepackage{enumerate}
\usepackage{nameref}
\usepackage{hyperref}
\usepackage{indentfirst}
\usepackage{float}%定义图片
\usepackage{multirow}
\usepackage{mathtools}
\usepackage{tabularx}
\usepackage{extarrows}
\usepackage{supertabular}
\usepackage{threeparttablex}
\usepackage{dcolumn}% Align table columns on decimal point
\usepackage{bm}% bold math
\usepackage{overpic}
\usepackage{booktabs}%
\usepackage{makecell}
\usepackage{booktabs} % For better table formatting

\graphicspath{{./}{figures/}}

\definecolor{navy}{RGB}{3, 37, 126}

\begin{document}

\title{GRB 250706B/C: \textit{Insight}-HXMT Discovery of a High-Luminosity Burst as a Candidate for Fallback-Regulated Accretion in the Prompt Emission}

\correspondingauthor{R.~Moradi, Shao-Lin Xiong}
\email{moradi@ihep.ac.cn, xiongsl@ihep.ac.cn}

\author[0009-0008-8053-2985]{Chen-Wei Wang}
\affil{State Key Laboratory of Particle Astrophysics, Institute of High Energy Physics, Chinese Academy of Sciences, Beijing 100049, China}
\affil{University of Chinese Academy of Sciences, Chinese Academy of Sciences, Beijing 100049, China}
\email{}

\author[0000-0002-2516-5894]{R.~Moradi}
\affil{State Key Laboratory of Particle Astrophysics, Institute of High Energy Physics, Chinese Academy of Sciences, Beijing 100049, China}
\email{rmoradi@ihep.ac.cn}

\author[0000-0002-4771-7653]{Shao-Lin Xiong}
\affil{State Key Laboratory of Particle Astrophysics, Institute of High Energy Physics, Chinese Academy of Sciences, Beijing 100049, China}
\email{xiongsl@ihep.ac.cn}

\author[0000-0001-5586-1017]{Shuang-Nan Zhang}
\affil{State Key Laboratory of Particle Astrophysics, Institute of High Energy Physics, Chinese Academy of Sciences, Beijing 100049, China}
\affil{University of Chinese Academy of Sciences, Chinese Academy of Sciences, Beijing 100049, China}
\email{}

\author[0009-0002-6411-8422]{Zheng-Hang Yu}
\affil{State Key Laboratory of Particle Astrophysics, Institute of High Energy Physics, Chinese Academy of Sciences, Beijing 100049, China}
\affil{University of Chinese Academy of Sciences, Chinese Academy of Sciences, Beijing 100049, China}
\email{}

\author[0009-0006-5506-5970]{Wen-Jun Tan}
\affil{State Key Laboratory of Particle Astrophysics, Institute of High Energy Physics, Chinese Academy of Sciences, Beijing 100049, China}
\affil{University of Chinese Academy of Sciences, Chinese Academy of Sciences, Beijing 100049, China}
\email{}

\author{Hao-Xuan Guo}
\affil{State Key Laboratory of Particle Astrophysics, Institute of High Energy Physics, Chinese Academy of Sciences, Beijing 100049, China}
\affil{Department of Nuclear Science and Technology, School of Energy and Power Engineering, Xi'an Jiaotong University, Xi'an, China}
\email{guohx@ihep.ac.cn}

\author{Xiao-Bo Li}
\affil{State Key Laboratory of Particle Astrophysics, Institute of High Energy Physics, Chinese Academy of Sciences, Beijing 100049, China}
\email{}

\author{Cheng-Kui Li}
\affil{State Key Laboratory of Particle Astrophysics, Institute of High Energy Physics, Chinese Academy of Sciences, Beijing 100049, China}
\email{}

\author{Jia-Cong Liu}
\affil{State Key Laboratory of Particle Astrophysics, Institute of High Energy Physics, Chinese Academy of Sciences, Beijing 100049, China}
\affil{University of Chinese Academy of Sciences, Chinese Academy of Sciences, Beijing 100049, China}
\email{}

\author{Xing-Hao Luo}
\affil{State Key Laboratory of Particle Astrophysics, Institute of High Energy Physics, Chinese Academy of Sciences, Beijing 100049, China}
\affil{Institute for Frontier in Astronomy and Astrophysics, Beijing Normal University, Beijing 102206, China, Department of Astronomy, Beijing Normal University, Beijing 100875, China}
\affil{Department of Astronomy, Beijing Normal University, Beijing 100875, China}
\email{202211998368@mail.bnu.edu.cn}

\author{Yang-Zhao Ren}
\affil{State Key Laboratory of Particle Astrophysics, Institute of High Energy Physics, Chinese Academy of Sciences, Beijing 100049, China}
\affil{School of Physical Science and Technology, Southwest Jiaotong University, Chengdu Sichuan, 611756, China}
\email{}

\author{Yue Wang}
\affil{State Key Laboratory of Particle Astrophysics, Institute of High Energy Physics, Chinese Academy of Sciences, Beijing 100049, China}
\affil{University of Chinese Academy of Sciences, Chinese Academy of Sciences, Beijing 100049, China}
\email{}

\author{Sheng-Lun Xie}
\affil{State Key Laboratory of Particle Astrophysics, Institute of High Energy Physics, Chinese Academy of Sciences, Beijing 100049, China}
\affil{Institute of Astrophysics, Central China Normal University, Wuhan 430079, China}
\email{}

\author{Wang-Chen Xue}
\affil{State Key Laboratory of Particle Astrophysics, Institute of High Energy Physics, Chinese Academy of Sciences, Beijing 100049, China}
\affil{University of Chinese Academy of Sciences, Chinese Academy of Sciences, Beijing 100049, China}
\email{}

\author{Yuan-Zao Xue}
\affil{State Key Laboratory of Particle Astrophysics, Institute of High Energy Physics, Chinese Academy of Sciences, Beijing 100049, China}
\affil{University of Chinese Academy of Sciences, Chinese Academy of Sciences, Beijing 100049, China}
\email{}

\author{Peng Zhang}
\affil{State Key Laboratory of Particle Astrophysics, Institute of High Energy Physics, Chinese Academy of Sciences, Beijing 100049, China}
\affil{College of Electronic and Information Engineering, Tongji University, Shanghai 201804, China}
\email{}

\author[0009-0001-7226-2355]{Chao Zheng}
\affil{State Key Laboratory of Particle Astrophysics, Institute of High Energy Physics, Chinese Academy of Sciences, Beijing 100049, China}
\affil{TIANFU Cosmic Ray Research Center, Chengdu, Sichuan, China}
\email{}

\begin{abstract}
Fallback accretion in collapsar models is often associated with underluminous gamma-ray bursts (GRBs), leading to the widespread view that fallback-fed engines may be intrinsically inefficient at producing high-luminosity events. In this Letter, we present GRB~250706B/C, a luminous long GRB observed by \textit{Insight}-HXMT that exhibits an unusual combination of extreme short-timescale variability and coherent large-scale temporal evolution. The prompt emission contains at least 79 resolved pulses and a minimum variability timescale of $\sim11$ ms. The pulse widths are nearly independent of photon energy and span a broad distribution with a median FWHM of $\sim0.30$ s, while the waiting times between adjacent pulses have a median of $\sim0.38$ s. The prompt-emission envelope exhibits a prolonged rise described by $F(t)\propto (t-t_0)^{0.47\pm0.01}$ followed by a rapid decline. Despite substantial pulse-to-pulse fluctuations, neither the pulse widths nor the waiting times show significant secular evolution during the main emission episode. 
These features indicate the coexistence of two distinct temporal components, including a slow evolving rising luminosity envelope and rapid stochastic variability. 
Such behavior is consistent with scenarios in which a time-dependent engine-feeding history regulates the large-scale emission while internal dissipation within the relativistic outflow produces the pulse structure. Within this context, GRB~250706B/C may represent a fallback-fed collapsar operating on a high-luminosity branch, suggesting that fallback itself does not necessarily limit the luminosity scale of GRBs.
\end{abstract}


\section{Introduction}

Long-duration gamma-ray bursts (GRBs) are widely associated with relativistic jets produced during the collapse of massive stars \citep[e.g.,][]{1993ApJ...405..273W,1999ApJ...524..262M}. In the collapsar scenario, delayed fallback of marginally bound stellar material after the initial explosion is a natural outcome of stellar collapse and has long been considered a way to prolong central engine activity beyond the initial infall phase \citep[e.g.,][]{1989ApJ...346..847C,2001ApJ...550..410M}. At the same time, fallback-fed engines are frequently discussed in connection with low-luminosity GRBs, whereas classical high-luminosity bursts are often attributed to an early phase of rapid hyperaccretion from the stellar core.

However, this contrast can be physically misleading. In the collapsar picture, the observed GRB luminosity is regulated primarily by how efficiently the inner accretion flow converts accreted mass into relativistic jet power, while fallback itself determines only how mass is supplied to the engine as a function of time at larger radii. In this sense, fallback regulates when mass arrives at the engine, not how efficiently that mass is ultimately converted into outflow energy. Early accretion models of GRBs have shown that fallback can, in principle, power luminous GRBs, provided that the returning material accretes efficiently rather than being largely expelled in winds \citep{2001ApJ...557..949N}. The commonly assumed link between fallback and low luminosity therefore reflects typical progenitor structure and disk-regime conditions, rather than a fundamental limitation of fallback-fed engines.

Motivated by these considerations, it is important to identify observational signatures that can distinguish between different central-engine feeding histories using the temporal properties of GRB prompt emission.
One possible framework that connects a time-dependent mass-supply history to prompt emission is the Accretion-Modulated Internal Shock \citep[AMIS;][]{2026ApJ...999..228M} model. In this framework, fallback is best viewed as one physically motivated realization of a time-modulated mass-delivery history, rather than as a unique identifier of low-luminosity outcomes. The secular evolution of the luminosity reflects the underlying mass-supply function, and if fallback leads to a sufficiently high accretion rate, the corresponding high luminosity becomes observable. In AMIS, the prompt emission variability arises from internal shocks among relativistic shells emanating from the central engine \citep[e.g.,][]{1994ApJ...430L..93R,1997ApJ...490...92K}, while the global behavior of the light-curve envelope traces the time-dependent engine feeding history. 

Nevertheless, disentangling these effects observationally requires a thorough characterization of individual burst temporal structures, including how variability timescales and pulse properties evolve along the sub-pulse sequence, and how the pattern of emission envelope is shaped. 
In practice, this has long been challenging for most GRBs, whose pulses are typically either too erratic to define a clear emission envelope, or too few to trace temporal trends in pulse characteristics.

In this work, we present the prompt emission observation of GRB 250706C, a high-luminosity burst with an unusualy dense sequence of narrow pulses. 
In Section \ref{sec:data}, a set of unique temporal features is revealed by detailed analysis, including a global rising trend, a rapid decay phase, and statistically stationary pulse widths and waiting times, which are, to some extent, consistent with the predictions of fallback accretion and provides hints into the nature of the central engine. 
Section \ref{sec:model} presents a brief discussion within the context of fallback regulated accretion, although this is not the only possible scenario. 
This work does not aim to establish AMIS as a universal central engine model for all GRBs. 
Instead, the study presents detailed observations of a special yet representative case, in which the prompt emission appears to reveal the imprint of both the central engine activity history and the jet dissipation processes.

\section{Observation, Data Reduction, and Analysis}\label{sec:data}

On July 6, 2025, at 16:45:26.900 UT (denoted as $T_0$), \textit{Insight}-HXMT/HE was triggered by the long bright burst GRB 250706C \citep{0706C_HXMT_GCN,2025arXiv251015816W}, which is also detected by Konus-Wind \citep{0706C_KW_GCN1,0706C_KW_GCN2}. 
A GeV counterpart was detected at $\sim$600\,s after $T_0$ \citep{LAT_GCN}. 
And another bright soft X-ray transient, named as GRB 250706B, was detected and localized by SVOM/ECLAIRs $\sim$1200\,s later than $T_0$. 
The localization from IPN \citep{0706C_IPN_GCN} suggests that GRB 250706B and GRB 250706C are not only temporally correlated but also spatially consistent, indicating the same burst. 
Actually, GRB 250706B is the X-ray afterglow of GRB 250706C, but it is so bright that it can trigger the SVOM/ECLAIRs independently. 
Therefore, we refer to this burst as GRB 250706B/C hereafter. 
The redshift is inferred to be $\sim0.942$ by VLT \citep{0706C_redshift_GCN}. 

The Hard X-ray Modulation Telescope (also named as \textit{Insight}-HXMT) is the first X-ray astronomical satellite of China, which was launched on June 15, 2017. 
Three sets of collimated telescopes are equipped on \textit{Insight}-HXMT, namely the High Energy X-ray telescope (HE), the Medium Energy X-ray telescope (ME), and the Low Energy X-ray telescope (LE) \citep{HXMT_zhang_2018,HXMT_zhang_2020,HXMT_Li_2020}. 
HE, the main force of GRB detection of \textit{Insight}-HXMT, consists of 18 NaI(Tl)/CsI(Na) phoswich scintillation detectors. 
With Pulse Shape Discrimination (PSD) technique, the signal from CsI and NaI can be distinguished. 
For GRB detection, $\gamma$-rays can penetrate the satellite platform and be detected by CsI detectors of HE (hereafter referred to as HXMT/HE-CsI) with a large effective area of $\sim$5000$\,\rm cm^2$ \citep{HE_calibration,HE_calibration_updated}. 

\begin{figure*}
\centering
\begin{tabular}{cc}
\begin{minipage}[b]{0.5\linewidth}
    \begin{overpic}[width=\textwidth]{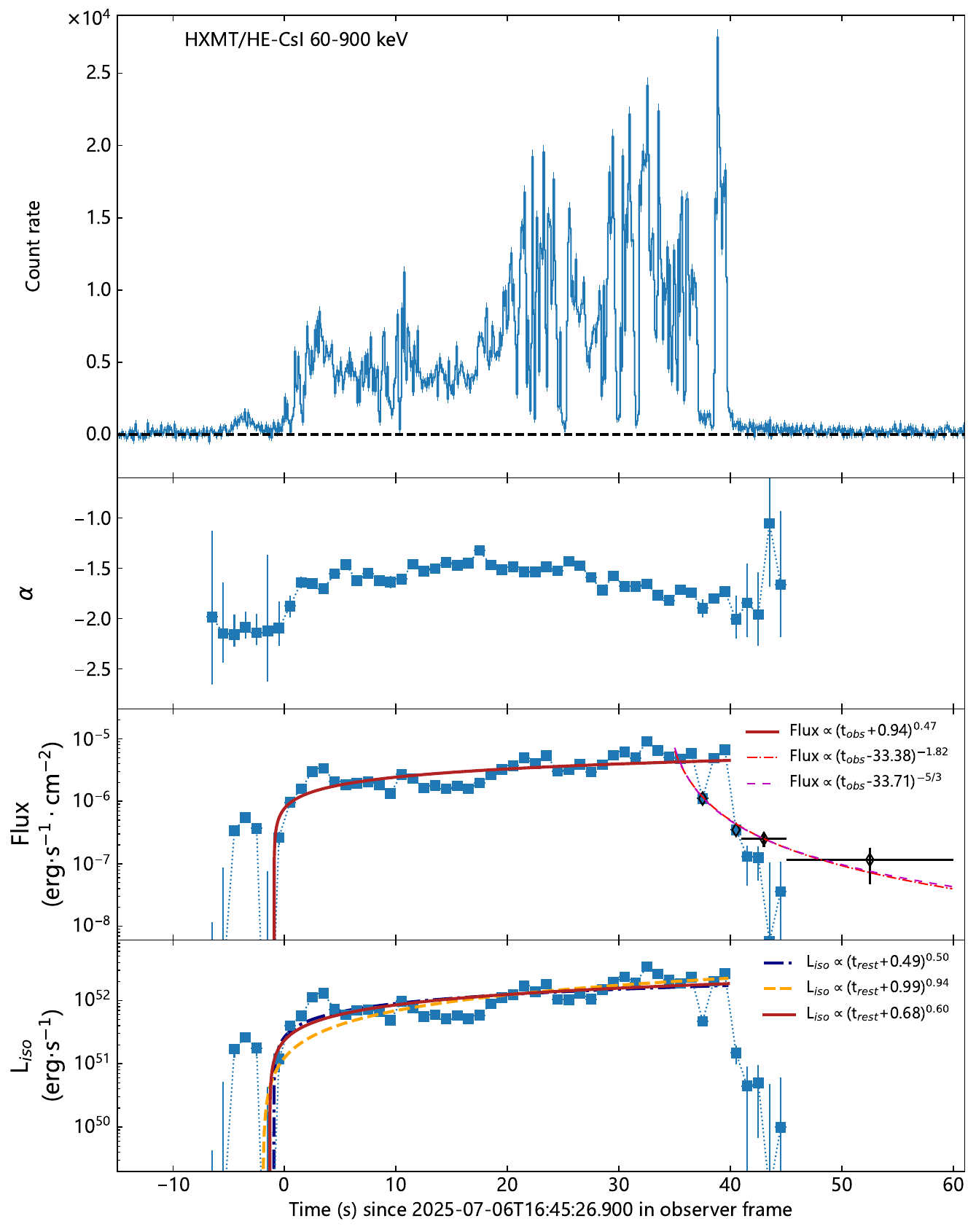}\put(0, 99){\bf a}\end{overpic}
\end{minipage}
\begin{minipage}[b]{0.44\linewidth}
        \begin{overpic}[width=\textwidth]{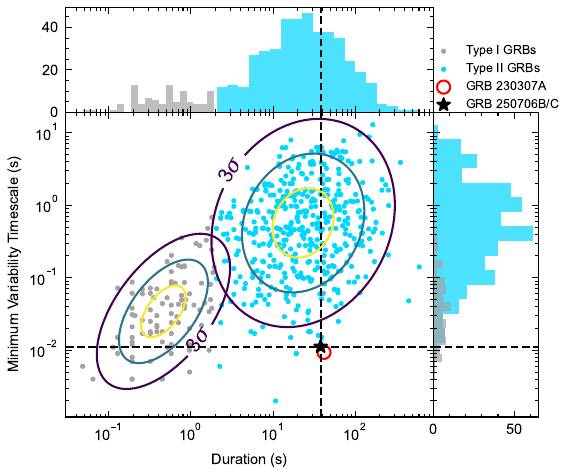}\put(-2, 80){\bf b}\end{overpic}
        \begin{overpic}[width=\textwidth]{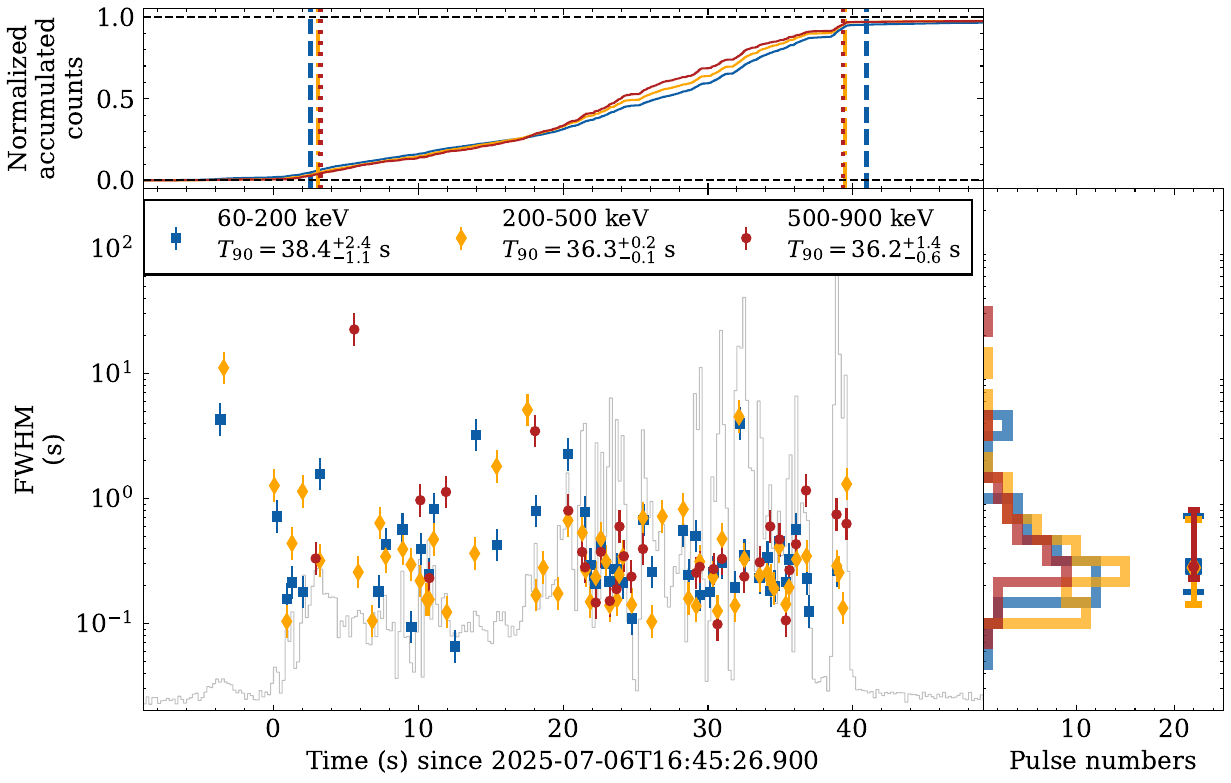}\put(-2, 65){\bf c}\end{overpic}
        % \begin{overpic}[width=\textwidth]{Fig_lc_sim.pdf}\put(-2, 80){\bf d}\end{overpic}
\end{minipage} \\
\end{tabular}
\caption{\textbf{The lightcurve of GRB 250706B/C.} 
(a), the time resolved spectra and the peak-finding result. 
(b), the GRB 250706B/C in $T_{90}$-MVT diagram. 
(c), the pulse width of GRB 250706B/C in three different energy range (i.e., 60-200 keV, 200-500 keV, 500-900 keV) as well as the duration of the emission in these three energy range. There is no significant broadening or narrowing of both the sub-pulses and the overall prompt emission as a function of photon energy, suggesting a central-engine–dominated emission process rather than a dissipation-dominated emission process. }
\label{fig:Fig_lc}
\end{figure*}

All HXMT/HE-CsI data are selected with pulse width from 90\,$\mu$s to 256\,$\mu$s. 
The energy range of CsI data used for spectral analysis is 120 to 600\,keV, while for temporal analysis the energy range is 60 to 900\,keV. 
The light curve of the HXMT/HE-CsI is depicted in the top panel of Fig.~\ref{fig:Fig_lc}a. 
Since the energy range of \textit{Insight}-HXMT/HE-CsI spectral analysis is not wide enough to constrain the $E_p$ of GRB 250706B/C \citep{0706C_KW_GCN2}, only a power-law (PL) model is adopted for spectral fitting, defined as $N(E)=AE^{-\alpha}$. The spectra are fitted using \texttt{Elisa} \citep{elisa2024}. 
The time-resolved spectral analysis results, including the photon index ($\alpha$) and flux measurements in the 120--600 keV band, are presented in Table~\ref{tab:spectral}. 

% One of the most notable features of GRB 250706B/C is the dense array of sharp, narrow pulses with extremely fast rising and steep decay. 
% The minimum variability time scale (MVT) of this burst is indeed very small, $\sim$0.011\,s, obtained using the Bayesian Block method following \cite{2025ApJ...979...73W}. 

One of the most notable features of GRB 250706B/C is the dense array of sharp, narrow pulses with extremely fast rising and steep decay. 
This rapid variability is quantitatively reflected in the minimum variability timescale (MVT), which, as determined by the Bayesian Block method \cite{2025ApJ...979...73W}, is remarkably small at $\sim$0.011\,s. 

As the MVT is usually energy-dependent (the MVT in the higher energy range is typically shorter), we applied \texttt{MEPSA} \citep{2015A&C....10...54G}, a flexible peak-finding algorithm, to identify pulses in four energy bands with light curve time resolution of 10\,ms (shorter than the MVT), including the full HXMT/HE-CsI band (60--900 keV) and three narrow bands (60--200 keV, 200--500 keV, and 500--900 keV). 
The analysis reveals at least 79 high signal-to-noise pulses in the full energy band, placing this burst among the most pulse-rich GRBs \citep{2024A&A...685A..34G}. 
The peak time, pulse width (described by FWHM) and peak rate, along with their uncertainties \citep{2023A&A...671A.112C}, are calculated for each pulse, which are listed in Table~\ref{tab:pulses}. 

Futuremore, the peak search result for the different energy bands suggests that the distribution of pulse width in GRB 250706B/C does not exhibit a significant dependence on energy, as shown in Fig.~\ref{fig:Fig_lc}c. 
Since the MVT is closely related to the pulse width, it can be inferred that the MVT of $\sim0.011$\,s should be relatively consistent across a broader range of energy bands. 
When comparing with other GRBs in the diagram of $T_{90}$--MVT, GRB 250706B/C becomes an outliner with MVT much smaller than that of the vast majority of long GRBs with similar durations, as portrayed in Fig.~\ref{fig:Fig_lc}b. 
It is interesting that the second brighest burst GRB 230307A, with more than 100 identifiable pulses \citep{2026JHEAp..4900456M}, falls into the same region with GRB 250706B/C in the $T_{90}$--MVT diagram. 
Moreover, the combination of its dense pulse structure and short MVT places GRB 250706B/C among the most extreme cases in the long-GRB population.

In addition to the MVT, another well-established property of GRB prompt emission is that energy dependence also applies to the overall pulse profile \citep[e.g.,][]{1995ApJ...448L.101F,1996ApJ...459..393N}, displaying a systematic ``softer-wider" trend (i.e., the duration in lower-energy bands is usually significantly longer than that in higher-energy bands). 
This phenomenon is generally interpreted as a consequence of radiative cooling, curvature effects, or other dissipation processes within relativistic outflows \citep[e.g.,][]{2004ApJ...614..827R,2005ApJ...632.1008Q}. 
In contrast, the observed constancy of the duration across all energy ranges of GRB 250706B/C, as shown in the top panel of Fig.~\ref{fig:Fig_lc}c, suggests a departure from these standard interpretations. 
Instead, such near energy-independence may be consistent with a prolonged central-engine regulation that is less sensitive to the details of the dissipation region. 
One plausible realization of this behavior is fallback accretion, whereby the sustained inflow of material onto the compact remnant drives long-lasting, quasi-thermal emission episodes with weak dependence on energy \citep[e.g.,][]{2008Sci...321..376K,2011ApJ...734...35C}. 
Therefore, the energy-independence of duration hints at a link between the prompt light-curve morphology and accretion-driven processes, providing a potential diagnostic of central engine activity in this GRB.

To investigate this, we performed power-law fits of the flux light curve of GRB 250706B/C, as well as the luminosity light curve, using a Bayesian Markov Chain Monte Carlo (MCMC) approach with the emcee package \citep{2013PASP..125..306F}

The interval from $T_0-1$\,s to $T_0+40$\,s of the flux light curve is well modeled with a power-law function of $F(t)\propto (t-t_0)^{0.47^{+0.01}_{-0.01}}$, indicating a clear rising trend in the prompt-emission envelope. 
This slope is consistent with the rising trend expected for the early stage of fallback accretion, but it is not a unique confirmation of that model, as other rising power-law scenarios, such as disk spreading or wind-fed accretion, can produce comparably shallow indices. 
The parameter $t_0$, the start time of the power-law rising trend, is also set as a free parameter, along with the normalization and slope, during the fitting. 
Moreover, the model function is integrated first within each time slice before being compared with the data by the $\chi^2$ during the fitting. 
This is because the horizontal error bars depicted in the flux light curve correspond to the integration intervals, rather than denoting temporal measurement uncertainties associated with individual data points. 
The precursor is not taken into consideration during the fitting as the significant quiescent period between the precursor and the main emission suggests that the precursor is not a part of the same rising power-law trend. 
Thus the prior on $t_0$ is set to be greater than -1, and the posterior of $t_0$ is -0.93$^{+0.06}_{-0.04}$. 
And the secular rising trend, as well as the energy-independence of the duration, together suggest that the central engine remains fueled by a time-dependent mass supply throughout the main emission episode.

Furthermore, we noticed that the flux light curve clearly shows a rapid decrease at the end of the prompt emission. 
Due to the very steep decay, the limited photon counts result in poor constraints on the spectra, preventing the flux light curve from maintaining high temporal resolution. 
Therefore, we combined the photons from 41 to 45 seconds into a single energy spectrum to reduce the uncertainty in flux measurement. 
Combined with another flux measurement taken with a longer exposure time after 45 seconds, the four black diamond data points in Fig.~\ref{fig:Fig_lc} were used to explore the decay slope of the final stage. 
The fitted slope is $\sim$-5/3, which is consistent with the typical prediction for a fallback-decay phase. 
However, the small number of late-time points and the steep decline make this result tentative. 
And the fitting result in Fig.~\ref{fig:Fig_lc}a suggests that the decay phase began before the rising trend fully ended. 
One possibility is that the high-flux intervals in the 38--40 s represent a flare if we take the decay phase as the early hard X-ray afterglow. 
Another plausible explanation may be that the rise persisted into the decay stage, suggesting that the fallback of stellar material is not strictly uniform or continuous. 
In any case, the rapid post-peak decline leads to the HXMT/HE band flux falling below the detection threshold, rather than indicating that the physical decay phase itself has an intrinsically short duration.

As for the isotropic-equivalent luminosity $L(t)$, the k-corrections based on the redshift are first applied and the time is also converted into the rest-frame time during the fitting with the same time interval selected as the fitting of the flux light curve. 
Moreover, three kinds of fitting with different strategies are conducted as follows to minimize the $t_0$ effect as much as possible:\\
(1) set normalization, slope and $t_0$ as free parameters, but the prior of $t_0$ is limited to greater than -1, which is the same prior setting as that used in the flux light curve fitting. 
In this scenario, the postior of slope is $0.94^{+0.02}_{-0.02}$ while $t_0$ exhibits convergence issues near the boundary of the prior distribution, which is $-0.98^{+0.03}_{-0.01}$. 
The $\chi^2/dof$ of this scenario is 2139/39.\\
(2) set normalization, slope and $t_0$ as free parameters, but the prior of slope is limited to less than 0.6. 
In this scenario, the postior of $t_0$ is $-0.68^{+0.06}_{-0.06}$ while slope exhibits convergence issues near the boundary of the prior distribution, whose median is -0.60. 
The $\chi^2/dof$ of this scenario is 2171/39.\\
(3) frozen slope as 0.5, only set normalization and $t_0$ as free parameters. 
In this scenario, the postior of $t_0$ is $-0.49^{+0.06}_{-0.05}$. 
The $\chi^2/dof$ of this scenario is 2231/40.\\
All the fitting results of these three scenrios are plotted in the Fig\,\ref{fig:Fig_lc}, converted to the observer time. 
Although the parameters are not well constrained, it still indicates that the temporal behavior of $L_{iso}$ in the rest frame behaves consistently with the flux in the observer frame, which exhibits a rising trend of $\sim t^{1/2}$.

Taking advantage of the large number of pulses in GRB 250706B/C, which enables a detailed study of the evolution of pulse properties, we also conducted a comprehensive analysis of the pulse peak rate, pulse width, and waiting time, defined as the time intervals between adjacent peaks, across the pulse sequence listed in Table\,\ref{tab:pulses}.

\begin{figure*}
    \centering
    \includegraphics[width=0.95\textwidth]{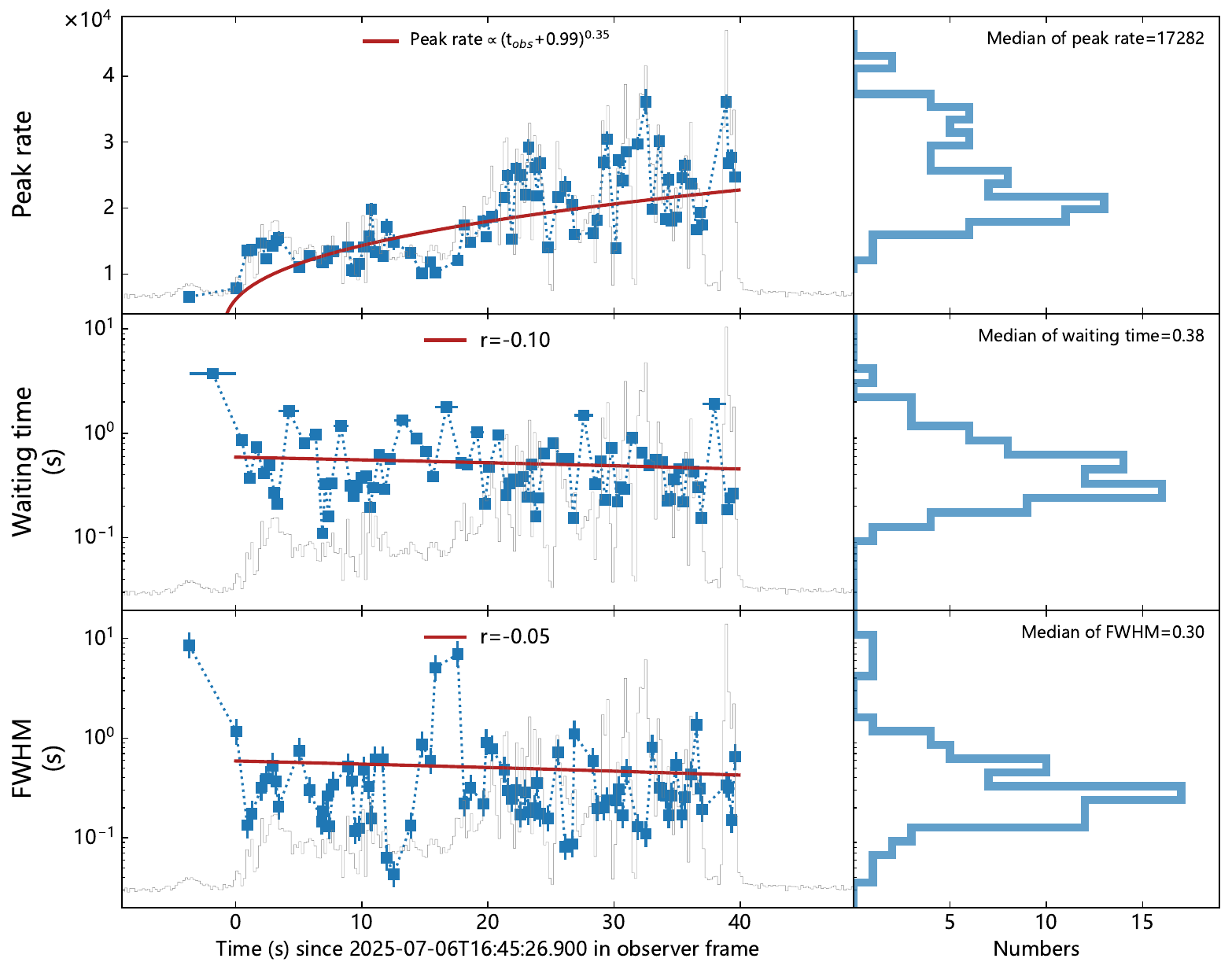}
    \caption{\textbf{Pulse distribution diagnostics for GRB 250706B/C.} The pulse width (FWHM) distribution, waiting-time distribution, and peak-rate distribution quantify the short-timescale variability of GRB 250706B/C. These histograms show broad scatter in pulse widths and inter-peak intervals, and a broad peak-rate distribution, indicating that the prompt variability is dominated by stochastic structure rather than a single preferred timescale or amplitude.}
    \label{fig:Fig_peak}
\end{figure*}

On the one hand, the distributions of the measured pulse properties, as shown in Figure\,\ref{fig:Fig_peak}, provide a stronger temporal characterization of the burst. 
For the 78 high-SNR peaks identified with \texttt{MEPSA}, the FWHM values span 0.04--8.49 s with a median of 0.30 s and a mean of 0.61 s, highlighting the broad scatter of pulse widths. 
While the waiting times range from 0.11 to 3.73 s with a median of 0.385 s and a mean of 0.56 s, suggesting that the burst does not favor a single preferred recurrence timescale. A lognormal fit to the FWHM distribution yields a median of 0.30 s and a scatter of 0.40 dex in $\log_{10}({\rm FWHM}/{\rm s})$. 
The peak-rate distribution is also well described by a lognormal shape, with a median of $1.7\times10^{4}$ counts s$^{-1}$ and a scatter of 0.16 dex, and it spans approximately $6.5\times10^{3}$--$3.6\times10^{4}$ counts s$^{-1}$, nearly an order of magnitude. 
The median pulse FWHM of 0.30 s is comparable to the internal dissipation timescale expected from emission radii of order $10^{14}$--$10^{15}$ cm for bulk Lorentz factors of order $100$--$300$, while the $\sim0.4$ s waiting-time scale may reflect the characteristic spacing of successive engine ejections or intermittent mass feeding rather than a single coherent oscillation. This estimate therefore depends on the assumed Lorentz factor and is intended as an illustrative, order-of-magnitude guide rather than a uniquely determined physical measurement. These physical estimates are illustrative rather than unique, and they underscore that the observed timescales are compatible with a rapidly variable inner engine.

\begin{figure}
    \centering
    \includegraphics[width=0.48\textwidth]{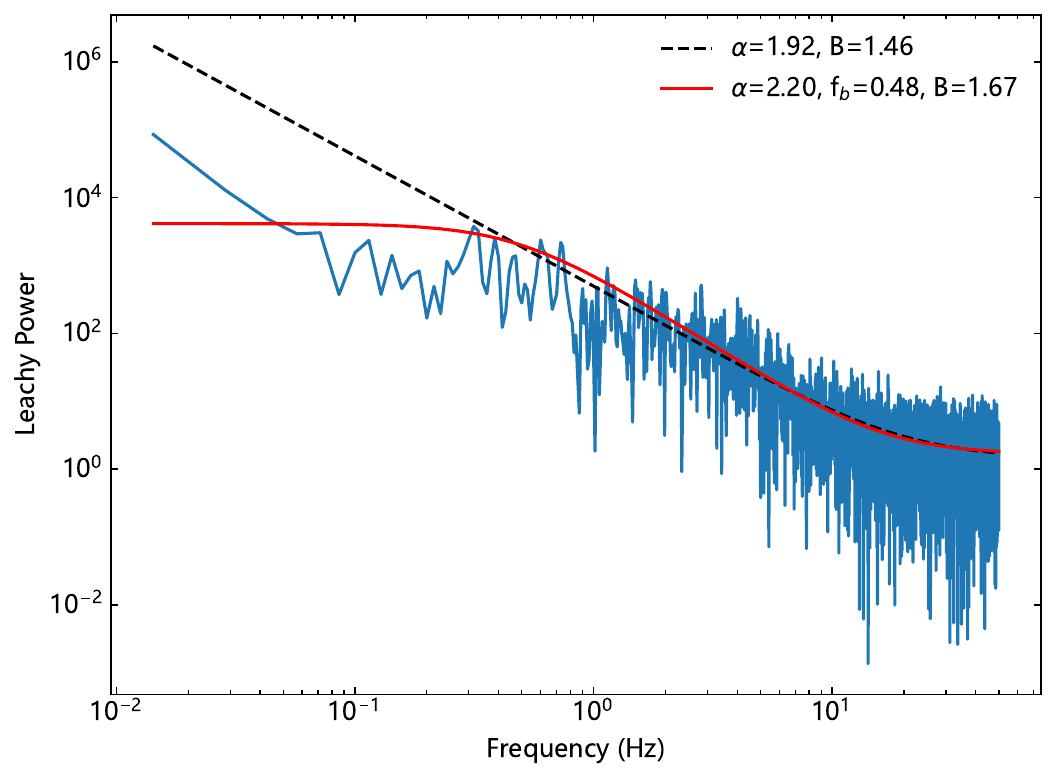}
    \caption{\textbf{Power spectral density of GRB 250706B/C with the light curve from $T_0$-10 s to $T_0$+60 s.} 
    The black dashed line and red line are thr fitting of a simple power-law plus the white-noise constant and bent power-law \citep{2016A&A...589A..98G}. 
    No narrow feature is detected significantly. }
    \label{fig:Fig_psd}
\end{figure}

On the other hand, the temporal evolution of these pulse properties reveals additional information beyond their static distributions.
The pulse peak rate shows an increasing trend, and can be fitted with a power-law model. 
As data points have normally distributed errors on both axes, the fitting is performed as a linear function in the log-log space by adopting the \cite{2005physics..11182D} likelihood. 
The best-fit posterior medians of the peak rate slope is $0.33^{+0.01}_{-0.01}$. 
However, the FWHM and waiting time do not show a significant increasing or decreasing trend over time. 
This can be further confirmed by the Pearson correlation coefficient (r), which is -0.10 and -0.05 for the waiting time and FWHM, respectively. 
Nevertheless, the pulse peak rate shows a modest negative correlation with FWHM ($r\approx-0.32$) and with waiting time ($r\approx-0.29$), while FWHM and waiting time are more strongly positively correlated ($r\approx0.61$). 
These trends are consistent with brighter pulses tending to be narrower and more closely spaced, but substantial scatter remains, indicating that the variability is not governed by a single simple scaling relation. 
These quantitative analysis reinforce that the pulse population is statistically broad and not dominated by a narrow, coherent timescale, which is also consistent with that no significant narrow-band periodic component is detected in the power spectral density, as portrayed in Fig~\ref{fig:Fig_psd}.

\begin{figure}
    \centering
    \includegraphics[width=0.48\textwidth]{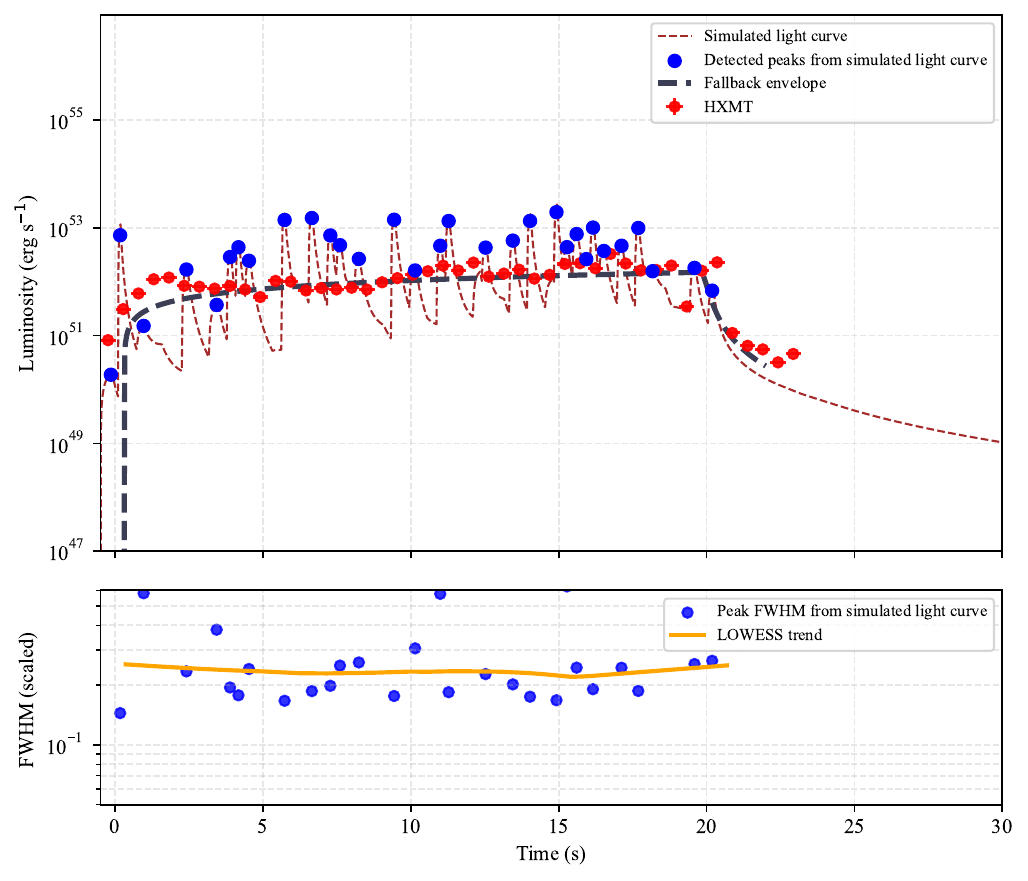}
    \caption{\textbf{Simulation of prompt emission variability under a mass-driven scenario for GRB 250706B/C.} The rising luminosity envelope, $L(t) \propto t^{0.5}$, results from shells ejected at roughly constant intervals with increasing mass, $m \propto t^{\beta}$, due to fallback accretion. Although the FWHM is not explicitly shown, the model predicts roughly constant characteristic pulse widths with no strong secular evolution across the burst; individual pulses fluctuate around this baseline, consistent with the observed ensemble behavior. The model separates the long-timescale envelope evolution from the short-timescale pulse variability, consistent with the statistical stationarity of the measured pulse widths and waiting times. This realization shows that a mass-driven process can reproduce the secular luminosity rise and the fine-scale temporal variability of the burst.}
    \label{fig:Fig_sim}
\end{figure}

\section{Discussion and Conclusions}
\label{sec:model}

% compare with other burst
\subsection{Comparsion with other burst}
Compared with other highly variable bright long GRBs such as GRB 130427A, GRB 250706B/C shows a distinctive temporal morphology, including a very short MVT, many sub-pulses, nearly energy-independent pulse widths as well as a temporal raising emission envelope.

% GRB 230307A
Among the high brightness events observed in the past, GRB 230307A is one of the few long bursts with a similarly large number of resolved pulses and has a nearly identical MVT to GRB 250706B/C \citep{2025NSRev..12E.401S}. 
However, the characteristic FWHM values and waiting time values of GRB 230307A are generally broader than those of GRB 250706B/C \citep{2026JHEAp..4900456M}. Moreover, the FWHM and waiting time of its sub-pulse sequence exhibit a gradually increasing trend, which is distinctly different from the nearly constant FWHM and waiting time observed in GRB 250706B/C \citep{2026JHEAp..4900456M}. 
Furthermore, although the narrow pulses of both GRB 230307A and GRB 250706/B combine to form an overall secular emission envelope, the overall profile of GRB 230307A exhibits a fast rise and a slow decay \citep{2024ApJ...977..155M,2025ApJ...985..239Y}. In contrast, GRB 250706B/C displays the opposite behavior, characterized by a long-term rise followed by a rapid decay. 
% GRB 130427A
GRB 130427A--the so-called ``Ordinary Monster", is also highly luminous and temporally structured, but shows stronger evidence for energy-dependent pulse broadening than GRB 250706B/C \citep{2014Sci...343...51P,2014Sci...343...48M,2026JHEAp..4900456M}.

% compare result
Therefore, GRB 250706B/C is significantly distinguished from other high-luminosity, high-variability gamma-ray bursts by its unique temporal behavior, placing GRB 250706B/C near the short-timescale end of the long-GRB population and making it a valuable benchmark for investigating a potentially new population of gamma-ray bursts.

\subsection{Possible Physical interpretation}

As discussed above, GRB~250706B/C exhibits a prompt luminosity envelope that rises over $\sim20$ s in the source frame and then declines, together with a pulse population characterized by a FWHM distribution centered near 0.30 s and a median waiting time near 0.4 s. Rapid pulse variability is superposed on this slowly evolving envelope, while both the pulse widths and waiting times remain statistically stationary throughout the main emission episode.

We consider one possible interpretation within the Accretion-Modulated Internal Shock (AMIS) framework \citep{2026ApJ...999..228M}. In this picture, the central engine ejects relativistic shells whose masses are modulated by the feeding rate, approximately $m_i\propto\dot{M}_{\rm fb}(t_i)\Delta t_{\rm ej}$ for quasi-regular ejection intervals $\Delta t_{\rm ej}$, while shell Lorentz factors fluctuate stochastically. The mass modulation governs the slowly varying kinetic-energy envelope, whereas Lorentz-factor contrasts determine where and when internal shocks occur and generate the pulse structure. Fallback accretion provides one realization of this time-dependent mass supply. The burst luminosity also depends on the efficiency with which accreted material is converted into jet power, parameterized by $\eta_{\rm acc}$. Further details of the AMIS framework and its numerical implementation are given in \citet{2026ApJ...999..228M}.

A simple estimate of the fallback radius is obtained by comparing the observed rise time with the free-fall time of material returning from radius $r$ to a black hole of mass $M_\bullet$, following the order-of-magnitude free-fall treatment used in collapsar fallback calculations \citep[e.g.,][]{2008MNRAS.388.1729K}:
\begin{equation}
{
\begin{aligned}
 t_{\rm ff}(r) \simeq \frac{\pi}{\sqrt{8}}\left(\frac{r^3}{G M_\bullet} \right)^{1/2} 
 &\simeq 1.4\,\rm s\left( \frac{r}{10^{9}\,\rm cm} \right)^{3/2} \\
 &\quad \times \left( \frac{M_\bullet}{5\,M_\odot} \right)^{-1/2}.
\end{aligned}
}
\end{equation}
Here $t_{\rm ff}(r)$ is the theoretical return time of material initially located at radius $r$. We denote by $t_0$ the onset time of the smooth rising component and by $t_p$ the approximate time at which this component reaches its broad maximum or turns over. These times are measured in the source frame. The relevant rise duration is therefore $\Delta t_p\equiv t_p-t_0$.

The characteristic fallback radius is obtained by setting $t_{\rm ff}(r_p)\simeq\Delta t_p$, which gives $r_p\simeq6.0\times10^9\,{\rm cm}\,(M_\bullet/5\,M_\odot)^{1/3}(\Delta t_p/20\,{\rm s})^{2/3}$. This estimate follows directly from the observed rise timescale and the free-fall relation above. Because the turnover is not sharply measured, $\Delta t_p\sim20$ s should be read as an order-of-magnitude source-frame timescale rather than as a precise fitted peak time.

Thus, a rise duration of order 20 s points to material returning from radii of several $10^9$ cm, characteristic of marginally bound envelope material in a collapsing Wolf--Rayet progenitor rather than the inner iron core. This order-of-magnitude estimate supports a fallback origin for the rising envelope and is consistent with material from the inner He-rich layers of a Wolf--Rayet progenitor \citep{1999ApJ...524..262M}.

For fallback-fed systems, the mass-return rate is expected to approach the ballistic scaling $\dot{M}_{\rm fb} \propto t^{-5/3}$ in the late-time limit \citep{1989ApJ...346..847C,2008MNRAS.388.1729K}. Hydrodynamic simulations show that this asymptotic regime follows an earlier transient phase sensitive to progenitor structure \citep{2001ApJ...550..410M}. We therefore use a phenomenological fallback envelope, rather than a unique hydrodynamic solution, to represent the observed shallow rise and the expected late-time decay:
\begin{equation}\label{eq:fb}
{
\dot{M}_{\rm fb}(t) \propto
\begin{cases}
(t-t_0)^\beta, & t_0<t\ll t_p,\\
(t-t_p)^{-5/3}, & t-t_p\gg\tau,
\end{cases}
}
\end{equation}
Here $t$ is the source-frame time, $t_0$ denotes the fallback-envelope onset, $t_p$ denotes the approximate turnover time, $\Delta t_p=t_p-t_0$ sets the rise duration, and $\tau$ is a short regularization timescale for the transition. The parameter $\beta$ describes the observed shallow rise; in the illustrative AMIS calculation we use $\beta=0.47$, consistent with the flux-envelope fit in Section~\ref{sec:data}. Equation~(\ref{eq:fb}) should be read as an asymptotic description: the envelope rises as $(t-t_0)^\beta$ and approaches the standard fallback decay $(t-t_p)^{-5/3}$ at late times \citep[see e.g.,][]{2001ApJ...550..410M,2013ApJ...767L..36W}. In the numerical AMIS implementation, the two branches are connected smoothly around $t_p$, and the overall shell-mass scale is set by the normalization $m_0$. The present data mainly constrain the rising branch; the available late-time flux points are too few and too uncertain to provide a robust independent fit to the $t^{-5/3}$ slope.

To illustrate this regime quantitatively, we performed forward AMIS simulations of a fallback-modulated engine. The simulation in Figure~\ref{fig:Fig_sim} adopts $N_{\rm sh}=70$ relativistic shells whose masses are modulated by Eq.~(\ref{eq:fb}) with normalization $m_0 = 8\times10^{27}$ g, $\beta=0.47$, and a source-frame turnover scale $\Delta t_p\sim20$ s. Shell Lorentz factors are drawn uniformly from $\Gamma\in[500,1100]$, while shell widths and ejection intervals remain narrowly distributed. The resulting outflow carries total kinetic energy $E_{\rm kin}\simeq2.9\times10^{53}$ erg, of which $E_{\rm diss}\simeq6.9\times10^{51}$ erg is dissipated in internal shocks ($\epsilon_{\rm IS}\simeq0.024$). For a fiducial jet conversion efficiency $\eta_{\rm acc}\sim0.1$, the prompt-phase accreted mass is $M_{\rm acc}\simeq1.6\,M_\odot$, implying $\dot{M}\simeq8\times10^{-2}\,M_\odot\,{\rm s}^{-1}$ over the same source-frame rise duration. The simulation is intended as an existence proof, not as a unique fit to the data.

The implied high accretion rate, places the system in the hyperaccreting regime. At such rates, neutrino cooling and magnetic stresses can reduce strong disk winds and allow efficient accretion onto the central object \citep{2001ApJ...557..949N}. In this picture, fallback regulates the mass-supply history, while the accretion efficiency determines the luminosity scale of the burst.

The temporal diagnostics of GRB~250706B/C give useful constraints on this interpretation. The prompt envelope rises approximately as $L(t)\propto t^{0.5}$, consistent with the rising branch of Eq.~(\ref{eq:fb}), while the pulse FWHM and waiting-time distributions remain statistically stationary despite strong pulse-to-pulse fluctuations. In a mass-driven internal-shock picture, pulse amplitudes tend to increase with the fallback supply, whereas the characteristic pulse timescales remain nearly unchanged. The late-time decline is broadly compatible with $t^{-5/3}$, but the current photon statistics make this tentative.

This interpretation is compatible with the AMIS framework, but it is not required by the observations. Magnetic reconnection, photospheric variability, and other rapid dissipation channels could also produce comparable pulse structure. We therefore do not claim that the present data identify internal shocks as the sole dissipation mechanism. The physical processes governing mass infall and accretion in collapsars are complex and remain under active study \citep{2022MNRAS.510.4962G,2025arXiv250208732I}. AMIS provides one physically motivated framework for connecting the observed temporal properties to the mass-feeding history of the central engine.

\subsection{Summary and caveats}

Thanks to the large effective area and high time resolution of \textit{Insight}-HXMT, we are able to conduct a detailed high-time-resolution study of GRB~250706B/C and reveal an unprecedented set of temporal properties. The prompt emission of GRB~250706B/C displays a secular luminosity rise approximately proportional to $t^{0.5}$, followed by a late-time decay that is approximately consistent with a $t^{-5/3}$ trend. Superposed on this envelope are nearly eighty narrow pulses, with characteristic pulse widths that remain approximately constant across energy bands and show no significant secular evolution throughout the burst. The waiting-time distribution also remains statistically stable despite substantial pulse-to-pulse fluctuations in amplitude and structure.

The temporal diagnostics presented in Figure~\ref{fig:Fig_peak} provide an observational characterization that is largely independent of any specific engine model. Figure~\ref{fig:Fig_peak} shows that the pulse-width, waiting-time, and peak-rate distributions are broad and approximately stochastic, with no evidence for a single preferred variability timescale. Figure~\ref{fig:Fig_psd} further shows that the pulse sequence is dominated by broad-band variability, with no statistically significant narrow periodicity or quasi-periodic oscillation features. Together, these results indicate that the prompt emission is governed by a complex, multi-timescale process rather than by a coherent clock-like mechanism. Any successful physical model must therefore explain both the slowly evolving luminosity envelope and the statistically stationary pulse population.

These observational results suggest that the prompt emission is governed by at least two distinct temporal components: a slowly evolving process that controls the overall luminosity envelope, and a rapidly varying process responsible for the fine temporal structure. The coexistence of these components, together with the absence of strong energy dependence in the pulse widths, indicates that the long-timescale evolution and short-timescale variability may originate from different physical mechanisms.

One possible interpretation of these temporal properties is provided by the Accretion-Modulated Internal Shock (AMIS) framework. In this picture, the global luminosity evolution is regulated by the time-dependent mass supply to the central engine, while the rapid variability is generated by internal dissipation within the relativistic outflow. The observed rise and decay of GRB~250706B/C are consistent with a fallback-regulated mass-supply history within the present uncertainties, whereas the pulse statistics can be associated with internal shock activity.

Within this interpretation, the high luminosity of GRB~250706B/C does not conflict with fallback regulation. Fallback determines when mass reaches the engine, while the accretion regime determines how efficiently that mass is converted into jet power. If the accretion flow remains efficient, as expected in hyperaccreting NDAF-like systems, fallback-fed engines can produce classical high-luminosity GRBs. Under less efficient conditions, the same fallback process may instead lead to underluminous events. The commonly discussed connection between fallback and low-luminosity GRBs may therefore reflect differences in accretion efficiency rather than differences in the fallback process itself.

While the data only constrain the viable physical interpretations rather than identifying a specific mechanism, the AMIS framework provides a physically natural explanation of the observed behavior, even though alternative scenarios involving different forms of central-engine activity or jet dissipation may also account for some of the observed properties.

Furthermore, the conclusions presented here are limited to the temporal properties of a single burst. 
To test whether the behavior observed in GRB~250706B/C typifies a long GRB subclass, and to investigate whether such features can be unambiguously attributed to a specific mechanism, a larger sample of bursts with comparable temporal complexity is needed for statistical analysis and comparative studies, an effort that relies on instruments with higher sensitivity and time resolution. 
Future polarization measurements of similar events with next-generation facilities such as POLAR-2 \citep{2025arXiv251109419H} and eXTP \citep{2025SCPMA..6819501Z} will also offer crucial clues into the underlying physical processes.

\section*{Acknowledgments}
We thanks S.~Kobayashi, Jesse Palmerio, Frederic Daigne, Dmitry Frederiks, Anna Ridnaia, Shu Zhang, Wen-Long Zhang and R. Maccary for useful discussions and suggestions. 
The authors used an AI-based tool solely for language polishing. 
R. Moradi acknowledges support from the Academy of Sciences Beijing Natural Science Foundation (IS24021) and the Institute of High Energy Physics, Chinese(E32984U810). 
We acknowledge the support by 
the National Key R\&D Program of China (2021YFA0718500), 
the National Natural Science Foundation of China (Grant No. 12494572, 
12273042,
12373047,
12333007,
),
the Strategic Priority Research Program of the Chinese Academy of Sciences (Grant Nos. 
XDA30050000, 
XDB0550300
) and China's Space Origins Exploration Program. 
This work made use of data from the \textit{Insight}-HXMT mission, funded by the CNSA and CAS. 
We are grateful to the development and operation teams of \textit{Insight}-HXMT. 

\bibliography{main}
\bibliographystyle{aasjournalv7}

\begin{center}
\begin{longtable*}{ccc}
\caption{Time-resolved Spectral Analysis Results. The table shows the time-resolved spectral analysis results with photon index ($\alpha$) and flux measurements in the 120-600 keV band. All errors represent 1$\sigma$ confidence intervals.\label{tab:spectral}} \\
\hline
Time interval (s) & $\alpha$ & Flux (erg cm$^{-2}$ s$^{-1}$) \\
\hline
\endfirsthead

\caption[]{(continued)} \\
\hline
Time interval (s) & $\alpha$ & Flux (erg cm$^{-2}$ s$^{-1}$) \\
\hline
\endhead

\hline
\endfoot

\hline
\endlastfoot
-6 $\sim$ -5 & $-2.15^{+0.50}_{-0.29}$ & $1.10^{+0.85}_{-1.09}\times10^{-7}$ \\
-5 $\sim$ -4 & $-2.16^{+0.20}_{-0.12}$ & $4.03^{+0.61}_{-0.63}\times10^{-7}$ \\
-4 $\sim$ -3 & $-2.09^{+0.16}_{-0.11}$ & $6.06^{+0.63}_{-0.59}\times10^{-7}$ \\
-3 $\sim$ -2 & $-2.14^{+0.19}_{-0.12}$ & $4.34^{+0.69}_{-0.67}\times10^{-7}$ \\
-2 $\sim$ -1 & $-2.12^{+0.76}_{-0.50}$ & $1.25^{+76.13}_{-1.24}\times10^{-9}$ \\
-1 $\sim$ 0 & $-2.10^{+0.26}_{-0.18}$ & $3.28^{+0.64}_{-0.64}\times10^{-7}$ \\
0 $\sim$ 1 & $-1.88^{+0.11}_{-0.11}$ & $1.05^{+0.08}_{-0.08}\times10^{-6}$ \\
1 $\sim$ 2 & $-1.64^{+0.05}_{-0.06}$ & $1.69^{+0.11}_{-0.11}\times10^{-6}$ \\
2 $\sim$ 3 & $-1.65^{+0.03}_{-0.04}$ & $3.12^{+0.12}_{-0.12}\times10^{-6}$ \\
3 $\sim$ 4 & $-1.70^{+0.03}_{-0.03}$ & $3.50^{+0.13}_{-0.12}\times10^{-6}$ \\
4 $\sim$ 5 & $-1.55^{+0.04}_{-0.04}$ & $2.19^{+0.11}_{-0.12}\times10^{-6}$ \\
5 $\sim$ 6 & $-1.46^{+0.04}_{-0.04}$ & $1.96^{+0.12}_{-0.13}\times10^{-6}$ \\
6 $\sim$ 7 & $-1.62^{+0.05}_{-0.05}$ & $2.05^{+0.11}_{-0.11}\times10^{-6}$ \\
7 $\sim$ 8 & $-1.55^{+0.04}_{-0.04}$ & $2.17^{+0.10}_{-0.12}\times10^{-6}$ \\
8 $\sim$ 9 & $-1.62^{+0.05}_{-0.05}$ & $1.97^{+0.10}_{-0.11}\times10^{-6}$ \\
9 $\sim$ 10 & $-1.64^{+0.06}_{-0.06}$ & $1.44^{+0.09}_{-0.11}\times10^{-6}$ \\
10 $\sim$ 11 & $-1.61^{+0.04}_{-0.04}$ & $2.81^{+0.11}_{-0.12}\times10^{-6}$ \\
11 $\sim$ 12 & $-1.46^{+0.03}_{-0.03}$ & $2.42^{+0.12}_{-0.12}\times10^{-6}$ \\
12 $\sim$ 13 & $-1.53^{+0.05}_{-0.05}$ & $1.75^{+0.11}_{-0.11}\times10^{-6}$ \\
13 $\sim$ 14 & $-1.50^{+0.04}_{-0.04}$ & $1.90^{+0.11}_{-0.11}\times10^{-6}$ \\
14 $\sim$ 15 & $-1.44^{+0.04}_{-0.04}$ & $1.71^{+0.11}_{-0.11}\times10^{-6}$ \\
15 $\sim$ 16 & $-1.47^{+0.04}_{-0.04}$ & $1.89^{+0.10}_{-0.12}\times10^{-6}$ \\
16 $\sim$ 17 & $-1.45^{+0.04}_{-0.04}$ & $1.71^{+0.10}_{-0.11}\times10^{-6}$ \\
17 $\sim$ 18 & $-1.32^{+0.03}_{-0.03}$ & $2.10^{+0.13}_{-0.14}\times10^{-6}$ \\
18 $\sim$ 19 & $-1.47^{+0.03}_{-0.03}$ & $2.83^{+0.12}_{-0.14}\times10^{-6}$ \\
19 $\sim$ 20 & $-1.51^{+0.03}_{-0.03}$ & $3.39^{+0.13}_{-0.14}\times10^{-6}$ \\
20 $\sim$ 21 & $-1.49^{+0.03}_{-0.02}$ & $3.84^{+0.14}_{-0.14}\times10^{-6}$ \\
21 $\sim$ 22 & $-1.54^{+0.02}_{-0.02}$ & $5.07^{+0.16}_{-0.14}\times10^{-6}$ \\
22 $\sim$ 23 & $-1.54^{+0.03}_{-0.03}$ & $4.13^{+0.15}_{-0.14}\times10^{-6}$ \\
23 $\sim$ 24 & $-1.49^{+0.02}_{-0.02}$ & $5.55^{+0.17}_{-0.17}\times10^{-6}$ \\
24 $\sim$ 25 & $-1.53^{+0.03}_{-0.03}$ & $3.16^{+0.12}_{-0.13}\times10^{-6}$ \\
25 $\sim$ 26 & $-1.43^{+0.03}_{-0.03}$ & $3.32^{+0.15}_{-0.14}\times10^{-6}$ \\
26 $\sim$ 27 & $-1.47^{+0.02}_{-0.02}$ & $4.07^{+0.15}_{-0.14}\times10^{-6}$ \\
27 $\sim$ 28 & $-1.59^{+0.03}_{-0.03}$ & $3.05^{+0.12}_{-0.12}\times10^{-6}$ \\
28 $\sim$ 29 & $-1.72^{+0.03}_{-0.03}$ & $3.97^{+0.13}_{-0.13}\times10^{-6}$ \\
29 $\sim$ 30 & $-1.57^{+0.02}_{-0.02}$ & $5.64^{+0.16}_{-0.17}\times10^{-6}$ \\
30 $\sim$ 31 & $-1.68^{+0.02}_{-0.02}$ & $6.34^{+0.15}_{-0.15}\times10^{-6}$ \\
31 $\sim$ 32 & $-1.68^{+0.02}_{-0.02}$ & $5.11^{+0.14}_{-0.15}\times10^{-6}$ \\
32 $\sim$ 33 & $-1.66^{+0.02}_{-0.02}$ & $9.26^{+0.18}_{-0.19}\times10^{-6}$ \\
33 $\sim$ 34 & $-1.77^{+0.02}_{-0.02}$ & $6.68^{+0.16}_{-0.15}\times10^{-6}$ \\
34 $\sim$ 35 & $-1.82^{+0.03}_{-0.02}$ & $5.22^{+0.12}_{-0.13}\times10^{-6}$ \\
35 $\sim$ 36 & $-1.71^{+0.03}_{-0.03}$ & $4.89^{+0.13}_{-0.14}\times10^{-6}$ \\
36 $\sim$ 37 & $-1.74^{+0.02}_{-0.02}$ & $6.02^{+0.12}_{-0.14}\times10^{-6}$ \\
37 $\sim$ 38 & $-1.90^{+0.09}_{-0.09}$ & $1.18^{+0.07}_{-0.08}\times10^{-6}$ \\
38 $\sim$ 39 & $-1.80^{+0.03}_{-0.03}$ & $5.05^{+0.13}_{-0.14}\times10^{-6}$ \\
39 $\sim$ 40 & $-1.73^{+0.02}_{-0.02}$ & $6.87^{+0.13}_{-0.16}\times10^{-6}$ \\
40 $\sim$ 41 & $-2.01^{+0.23}_{-0.20}$ & $4.15^{+0.64}_{-0.67}\times10^{-7}$ \\
41 $\sim$ 42 & $-1.85^{+0.40}_{-0.34}$ & $2.17^{+0.64}_{-0.86}\times10^{-7}$ \\
42 $\sim$ 43 & $-1.96^{+0.42}_{-0.31}$ & $1.98^{+0.66}_{-0.72}\times10^{-7}$ \\
43 $\sim$ 44 & $-1.05^{+0.59}_{-0.63}$ & $5.92^{+9.97}_{-5.34}\times10^{-8}$ \\
44 $\sim$ 45 & $-1.66^{+0.73}_{-0.52}$ & $1.38^{+0.72}_{-1.02}\times10^{-7}$ \\
\hline
41 $\sim$ 45 & $-1.87^{+0.32}_{-0.32}$ & $2.49^{+0.64}_{-0.67}\times10^{-8}$ \\
45 $\sim$ 60 & $-2.05^{+0.39}_{-0.37}$ & $1.16^{+0.69}_{-0.65}\times10^{-8}$ \\
\end{longtable*}
%\tablecomments{}
\end{center}

\begin{center}
\begin{longtable*}{ccc}
\caption{Pulse Characteristics for GRB 250706C identified with \texttt{MEPSA} \citep{2015A&C....10...54G}. 
FWHM uncertainties are asymmetric 1$\sigma$ errors.\label{tab:pulses}} \\
\hline
Peak Time & Peak Rate & FWHM \\
(s) & (counts s$^{-1}$) & (s) \\
\hline
\endfirsthead

\caption[]{(continued)} \\
\hline
Peak Time & Peak Rate & FWHM \\
(s) & (counts s$^{-1}$) & (s) \\
\hline
\endhead

\hline
\endfoot

\hline
\endlastfoot
-3.675 & 6546$\pm$124 & 8.49$^{+2.96}_{-2.20}$\\
0.055 & 7863$\pm$280 & 1.16$^{+0.41}_{-0.30}$\\
0.915 & 13567$\pm$823 & 0.13$^{+0.05}_{-0.03}$\\
1.285 & 13672$\pm$826 & 0.18$^{+0.06}_{-0.05}$\\
2.020 & 14714$\pm$700 & 0.32$^{+0.11}_{-0.08}$\\
2.430 & 12291$\pm$495 & 0.39$^{+0.14}_{-0.10}$\\
2.925 & 14270$\pm$487 & 0.53$^{+0.18}_{-0.14}$\\
3.195 & 15044$\pm$613 & 0.37$^{+0.13}_{-0.09}$\\
3.405 & 15546$\pm$881 & 0.21$^{+0.07}_{-0.05}$\\
5.045 & 11090$\pm$429 & 0.74$^{+0.26}_{-0.19}$\\
5.850 & 12808$\pm$653 & 0.30$^{+0.11}_{-0.08}$\\
6.825 & 11846$\pm$769 & 0.15$^{+0.05}_{-0.04}$\\
6.935 & 11803$\pm$543 & 0.18$^{+0.06}_{-0.05}$\\
7.260 & 12379$\pm$642 & 0.26$^{+0.09}_{-0.07}$\\
7.420 & 13522$\pm$671 & 0.13$^{+0.05}_{-0.03}$\\
7.750 & 13419$\pm$668 & 0.34$^{+0.12}_{-0.09}$\\
8.925 & 14096$\pm$593 & 0.52$^{+0.18}_{-0.13}$\\
9.240 & 10588$\pm$460 & 0.37$^{+0.13}_{-0.10}$\\
9.490 & 10442$\pm$589 & 0.12$^{+0.04}_{-0.03}$\\
9.795 & 11527$\pm$759 & 0.12$^{+0.04}_{-0.03}$\\
10.170 & 14119$\pm$531 & 0.48$^{+0.17}_{-0.13}$\\
10.560 & 15749$\pm$724 & 0.33$^{+0.11}_{-0.08}$\\
10.755 & 19802$\pm$995 & 0.16$^{+0.05}_{-0.04}$\\
11.055 & 13293$\pm$470 & 0.61$^{+0.21}_{-0.16}$\\
11.680 & 12739$\pm$504 & 0.61$^{+0.21}_{-0.16}$\\
11.970 & 17128$\pm$1308 & 0.06$^{+0.02}_{-0.02}$\\
12.540 & 14834$\pm$1217 & 0.04$^{+0.02}_{-0.01}$\\
13.875 & 13214$\pm$812 & 0.13$^{+0.05}_{-0.03}$\\
14.770 & 10080$\pm$278 & 0.87$^{+0.30}_{-0.22}$\\
15.440 & 11845$\pm$486 & 0.59$^{+0.21}_{-0.15}$\\
15.825 & 10212$\pm$190 & 5.06$^{+1.77}_{-1.31}$\\
17.600 & 12085$\pm$167 & 6.95$^{+2.43}_{-1.80}$\\
18.120 & 17437$\pm$762 & 0.22$^{+0.08}_{-0.06}$\\
18.620 & 14798$\pm$702 & 0.32$^{+0.11}_{-0.08}$\\
19.645 & 18033$\pm$949 & 0.22$^{+0.08}_{-0.06}$\\
19.855 & 15638$\pm$360 & 0.90$^{+0.32}_{-0.23}$\\
20.330 & 18657$\pm$610 & 0.78$^{+0.27}_{-0.20}$\\
21.295 & 21592$\pm$734 & 0.48$^{+0.17}_{-0.12}$\\
21.550 & 24943$\pm$911 & 0.30$^{+0.10}_{-0.08}$\\
21.880 & 15314$\pm$714 & 0.25$^{+0.09}_{-0.06}$\\
22.240 & 26004$\pm$931 & 0.30$^{+0.11}_{-0.08}$\\
22.585 & 24991$\pm$1117 & 0.17$^{+0.06}_{-0.04}$\\
22.970 & 22055$\pm$857 & 0.28$^{+0.10}_{-0.07}$\\
23.215 & 29234$\pm$1209 & 0.18$^{+0.06}_{-0.05}$\\
23.715 & 26131$\pm$1143 & 0.20$^{+0.07}_{-0.05}$\\
23.875 & 21863$\pm$739 & 0.36$^{+0.12}_{-0.09}$\\
24.115 & 26823$\pm$1158 & 0.18$^{+0.06}_{-0.05}$\\
24.755 & 14038$\pm$837 & 0.16$^{+0.05}_{-0.04}$\\
25.555 & 21692$\pm$601 & 0.72$^{+0.25}_{-0.19}$\\
26.120 & 23299$\pm$1526 & 0.08$^{+0.03}_{-0.02}$\\
26.690 & 20506$\pm$1432 & 0.09$^{+0.03}_{-0.02}$\\
26.845 & 16007$\pm$447 & 1.10$^{+0.38}_{-0.29}$\\
28.330 & 16215$\pm$569 & 0.59$^{+0.20}_{-0.15}$\\
28.655 & 18163$\pm$952 & 0.20$^{+0.07}_{-0.05}$\\
29.195 & 26918$\pm$1160 & 0.20$^{+0.07}_{-0.05}$\\
29.425 & 30427$\pm$1233 & 0.24$^{+0.08}_{-0.06}$\\
30.150 & 13932$\pm$681 & 0.24$^{+0.08}_{-0.06}$\\
30.370 & 27269$\pm$953 & 0.31$^{+0.11}_{-0.08}$\\
30.675 & 24198$\pm$1099 & 0.17$^{+0.06}_{-0.04}$\\
30.965 & 28546$\pm$844 & 0.46$^{+0.16}_{-0.12}$\\
31.870 & 29739$\pm$995 & 0.13$^{+0.04}_{-0.03}$\\
32.520 & 36125$\pm$1900 & 0.11$^{+0.04}_{-0.03}$\\
33.015 & 19838$\pm$575 & 0.81$^{+0.28}_{-0.21}$\\
33.570 & 30162$\pm$1002 & 0.32$^{+0.11}_{-0.08}$\\
34.100 & 18348$\pm$782 & 0.27$^{+0.09}_{-0.07}$\\
34.325 & 24270$\pm$1101 & 0.17$^{+0.06}_{-0.04}$\\
34.560 & 18128$\pm$777 & 0.26$^{+0.09}_{-0.07}$\\
34.920 & 18619$\pm$610 & 0.54$^{+0.19}_{-0.14}$\\
35.375 & 24608$\pm$1109 & 0.17$^{+0.06}_{-0.04}$\\
35.595 & 26478$\pm$1150 & 0.25$^{+0.09}_{-0.07}$\\
36.095 & 23727$\pm$770 & 0.43$^{+0.15}_{-0.11}$\\
36.525 & 16707$\pm$323 & 1.36$^{+0.47}_{-0.35}$\\
36.830 & 19345$\pm$803 & 0.31$^{+0.11}_{-0.08}$\\
36.985 & 17452$\pm$934 & 0.19$^{+0.07}_{-0.05}$\\
38.890 & 36116$\pm$1097 & 0.34$^{+0.12}_{-0.09}$\\
39.075 & 26766$\pm$667 & 0.32$^{+0.11}_{-0.08}$\\
39.315 & 27684$\pm$1176 & 0.15$^{+0.05}_{-0.04}$\\
39.580 & 24689$\pm$593 & 0.65$^{+0.23}_{-0.17}$\\
\end{longtable*}
%\tablecomments{}
\end{center}

\end{document}